\documentstyle[10pt,fullpage,psfig,bbbold,twocolum]{article}

\renewcommand{\baselinestretch}{1.281}
\renewcommand{\baselinestretch}{0.99}
\def\slantfrac#1#2{\hbox{$\,^{#1}\!/_{#2}$}}
\def\dfrac#1#2{{\displaystyle#1\over\displaystyle#2}}
\def\lqq{\lq\lq}

\begin{document}

\title{\LARGE Synthetic Doppler maps of gaseous flows in IP Peg}
\author{\Large O.A.Kuznetsov$^1$,
D.V.Bisikalo$^2$, A.A.Boyarchuk$^2$,\\[0.3cm]
\Large T.S.Khruzina$^3$, A.M.Cherepashchuk$^3$\\[0.9cm]
$^1$ {\it Keldysh Institute of Applied Mathematics, Moscow}\\
{\sf kuznecov@spp.keldysh.ru}\\[0.3cm]
$^2$ {\it Institute of Astronomy of the Russian Acad. of Science,
Moscow}\\
{\sf bisikalo@inasan.rssi.ru; aboyar@inasan.rssi.ru}\\[0.3cm]
$^3$ {\it Sternberg Astronomical Institute, Moscow}\\
{\sf kts@sai.msu.su; cher@sai.msu.su}\\[0.3cm]}
\date{}
\maketitle

\begin{abstract}

We present synthetic Doppler maps of gaseous flows in binary IP
Peg based on the results of 3D gasdynamical simulations. Using
of gasdynamical calculations alongside with the Doppler
tomography technique permits us to identify the main features of
the flow on the Doppler maps without solution of an ill-posed
inverse problem. Comparison of synthetic tomograms with
observations shows that in quiescence there are two zones of
high emission: a shock wave on the edge of the stream from $L_1$
caused by the interaction of the gas of circumbinary envelope
and the stream, and the dense region in a apoastron of
quasi-elliptical accretion disk. A single arm of the spiral
shock and the stream itself give a minor input to the total
brightness. During outburst the accretion disk dominates, and
the most emitive regions are the two arms of the spiral shock.

\end{abstract}

\section{Introduction}

Traditional observations of binary systems are carried out using
photometric and spectroscopic methodics. The former gives the
time dependence of brightness in a specific band $I(t)$ and the
latter can give the time dependence of wavelength of some
Doppler-shifted line $\lambda(t)$. Given ephemeris $\varphi(t)$
is known, the dependencies $I(t)$ and $\lambda(t)$ can be
converted by virtue of Doppler formula to the light curve
$I(\varphi)$ and the phase dependency of radial velocity
$V_R(\varphi)$.

During the last ten years the observations of binary systems in
the form of trailed spectrograms for some emission line
$I(\lambda,t)$ (or in other terms $I(V_R,\varphi)$) become
widely used. A method of Doppler tomography (Marsh \& Horne
1988$^{\cite{MarshHorne88}}$) is suited to analyze the trailed
spectrograms. Using this method one can obtain a map of
luminosity in the 2D velocity space from the orbital variability
of emission lines intensity. The Doppler tomogram is constructed
as a conversion of time resolved (i.e. phase-folded) line
profiles into the map on ($V_x$, $V_y$) plane. To convert the
distribution $I(V_R,\varphi)$ into the Doppler map $I(V_x,V_y)$
we should use the expression for radial velocity as a projection
of velocity vector on the line of sight, i.e.
$V_R=-V_x\cos(2\pi\varphi)+V_y\sin(2\pi\varphi)$ (here we assume
that $V_z\sim0$, the minus sign before $V_x$ is taken for
consistency with the coordinate system), and solve an inverse
problem that is described by integral equation (see Appendix A
of Marsh \& Horne 1988$^{\cite{MarshHorne88}}$):

\begin{equation}
\begin{array}{ll}
I(V_R,\phi)=\displaystyle\int\int I(V_x,V_y)\\
~\\
\qquad\times
g(V_R+V_x\cos(2\pi\phi)-V_y\sin(2\pi\phi))\\
~\\
\qquad\times
dV_x dV_y\,,
\end{array}
\label{eq_appendix}
\end{equation}
where $g(V)$ is a normalized local line profile (e.g., the
Dirac $\delta$-function), and the limits of integration are from
$-\infty$ to $+\infty$. This inverse problem is ill-posed and
the special regularization is necessary for its solving (e.g.
Maximum Entropy Method, Narayan \& Nityananda
1986$^{\cite{NarayanNit86}}$, Marsh \& Horne
1988$^{\cite{MarshHorne88}}$; Fourier Filtered Back Projection,
Robinson, Marsh \& Smak 1993$^{\cite{Robinson93}}$; Fast Maximum
Entropy Method, Spruit 1998$^{\cite{Spruit98}}$; etc., see also
Frieden 1979$^{\cite{Frieden79}}$).

As a result we obtain a map of distribution of specific line
intensity in velocity space. This map is easier to interpret
than original line profiles, moreover, the tomogram can show (or
at least gives a hint to) some features of flow structure. In
particular, the double-peaked line profiles corresponding to
circular motion of the gas (e.g. in accretion disk) become a
diffuse ring-shaped region in this map. Resuming, we can say
that components of binary system can be resolved in velocity
space while they can not be spatially resolved through direct
observations, so the Doppler tomography technique is a rather
power tool for studying of binary systems.

Unfortunately, the reconstruction of the spatial distribution of
intensity on the basis of a Doppler map is an ambiguous problem
since the points located far from each other may have equal
radial velocities and deposit to the same pixel on the Doppler
map. So the transformation $I(V_x,V_y)\to I(x,y)$ is impossible
without some {\it a priori} assumptions on the velocity field.

The situation changes drastically when one uses gasdynamical
calculations alongside with Doppler tomography technique. In
this case we don't need to cope with the inverse problem since
the task is solved directly: $\rho(x,y)~\&~T(x,y)\to I(x,y)$ and
$I(x,y)~\&~V_x(x,y)~\&~V_y(x,y)\to I(V_x,V_y) \to
I(V_R,\varphi)$. Difficulties can arise when converting the
spatial distributions of density and temperature $\rho(x,y)$,
$T(x,y)$ into the distribution of luminosity of specific
emission line $I(x,y)$. For optically thick lines the formation
of line profile should be described by radiation transfer
equations (see, e.g., Horne \& Marsh
1986$^{\cite{HorneMarsh86}}$), therefore to produce synthetic
Doppler maps we assume that the matter is optically thin and the
intensity of considered recombination line is as $I \sim \rho^2
T^{1/2}$ (Ferland 1980$^{\cite{Ferland80}}$; Richards \& Ratliff
1998$^{\cite{Richards98}}$).

\section{Binary system parameters}

The variable star IP Peg was discovered by Lipovetskij and
Stepanyan (1981$^{\cite{Lipovetskij}}$). It was found to be an
eclipsing dwarf nova (orbital period $3^{\rm h}\!.79$) with a
deep eclipse and a hump on the light curve by Goransky et al.
(1985$^{\cite{Goransky}}$).

A binary system can be completely described by the following set
of parameters: orbital period $P$, masses of the components
$M_1$ and $M_2$, separation $A$, and inclination angle $i$.
Among all parameters only $P$ is determined immediately from
observations while parameters $M_1$, $M_2$, $A$, $i$ (or in
another statement $M$, $q$, $A$, $i$, where $M=M_1+M_2$ -- total
mass, $q=M_2/M_1$ -- mass ratio) should be calculated from
observational data. These parameters are connected by third
Kepler's law $\Omega^2A^3=GM$ (here $\Omega=2\pi/P$ -- angular
velocity of the orbital motion; $G$ -- gravitational constant),
so we need only three additional relations.

Several ways of determination of relationships between the
parameters of the binary on the basis of observational data are
known to exist:

(i) using semi-amplitudes of radial velocity

$$
K_1=A\cdot\Omega\cdot\sin i\cdot\frac{M_2}{M}=A\cdot\Omega\cdot\sin i\cdot\frac{q}{1+q}
\,,
$$

$$
K_2=A\cdot\Omega\cdot\sin i\cdot\frac{M_1}{M}=A\cdot\Omega\cdot\sin i\cdot\frac{1}{1+q}
$$
for mass-losing star and accretor, correspondingly, we can
calculate $q=K_1/K_2$, and set a relation between $A$ and
$i$ ($\Omega$ is considered to be already known);

(ii) using the width of the white dwarf eclipse (for eclipsing
binaries only) we can set a relation between $q$ and $i$.

Knowing observable entities $K_1$, $K_2$, and $\Delta\phi$
permits us to determine all parameters of the binary system.
Note, that alternate methods of setting of relations between
system's parameters also exist. For example, knowing of
rotational broadening of absorbtion lines
$V_{rot}=R_{RL}(q,A)\Omega\sin i$, (where $R_{RL}$ -- effective
radius of the Roche lobe) permits set an additional relation
between $A$, $q$ and $i$. Usage of extra relations can be
exploited for checking of input
parameters.\footnote{Determination of system's parameters from
width of `hot spot' eclipse (see, e.g., Wood \& Crawford
1986$^{\cite{WoodCrawford86}}$; Smak 1996$^{\cite{Smak96}}$)
is not considered here since we suggest an essentially
different model of the flow structure in semidetached binaries.
We also don't rely on the mass-radius relation since it is
determined primarily for single stars.}

Semi-amplitudes of radial velocity for white dwarf were
determined in Wood \& Crawford
(1986$^{\cite{WoodCrawford86}}$); Marsh 
(1988$^{\cite{Marsh88}}$); Marsh \& Horne 
(1990$^{\cite{MarshHorne90}}$): Marsh (1988$^{\cite{Marsh88}}$) 
obtained the value of $K_1=175\pm15$ km~s$^{-1}$ (but using 
Shafter's method (1983$^{\cite{Shafter83}}$) he obtained a lower 
value of $K_1=164$ km~s$^{-1}$),  Wood \& Crawford 
(1986$^{\cite{WoodCrawford86}}$) obtained the value of $K_1=141$
km~s$^{-1}$. Semi-amplitudes of radial velocity for mass-losing
star were determined in Martin et al.
(1987$^{\cite{Martin87}}$, 1989$^{\cite{Martin89}}$); Marsh
(1988$^{\cite{Marsh88}}$); Beekman et al.
(2000$^{\cite{Beekman2000}}$): Marsh (1988$^{\cite{Marsh88}}$)
obtained the value of $K_2=305\pm15$ km~s$^{-1}$, Martin et al.
(1987$^{\cite{Martin87}}$, 1989$^{\cite{Martin89}}$) -- the
value of $K_2=288-298$ km~s$^{-1}$, and Beekman et al.
(2000$^{\cite{Beekman2000}}$) -- the value of $K_2=331\pm5.8$
km~s$^{-1}$. We adopt the values of semi-amplitudes of radial
velocity according to Wood \& Crawford
(1986$^{\cite{WoodCrawford86}}$) and Marsh \& Horne
(1990$^{\cite{MarshHorne90}}$): $K_1=148$ km~s$^{-1}$, $K_2=301$
km~s$^{-1}$.

\begin{figure}[t]
\centerline{\hbox{\psfig{figure=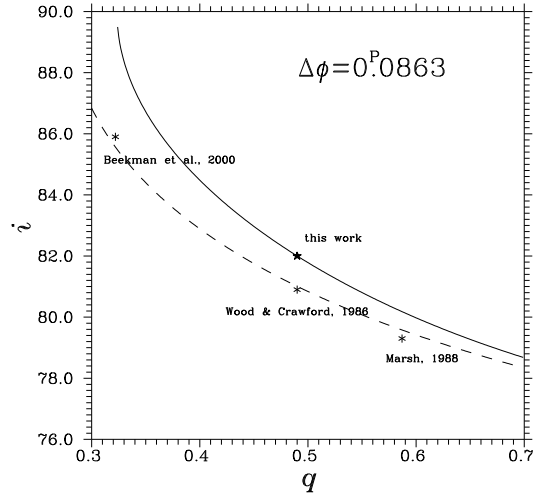,width=7cm}}}
\caption{\footnotesize Dependence between the orbit inclination
$i$ and the mass ratio $q$ for the value of width of the white
dwarf eclipse $\Delta\phi=0^{\rm P}\!\!.0863$.  A dashed line
corresponds to an approximate formula for the Roche lobe radius
(eq.~(\ref{eq2}), see Pacy\'nski 1971$^{\protect\cite{pacyn71}}$). A
solid line corresponds to the exact formula for the Roche lobe
sizes in $Y$ and $Z$ directions. Besides, we put values $i$ and
$q$ adopted by Wood \& Crawford
(1986$^{\protect\cite{WoodCrawford86}}$); Marsh
(1988$^{\protect\cite{Marsh88}}$); Beekman et al.
(2000$^{\protect\cite{Beekman2000}}$) as well as values adopted in our
work: $q=0.49$, $i=82^\circ$.} \end{figure}

After the values of $K_1$ and $K_2$ are adopted we can exploit
the relation between system's parameters resulting from the
value of width of the white dwarf eclipse. This value is known
with sufficient accuracy: Wood \& Crawford
(1986$^{\cite{WoodCrawford86}}$) obtained $\Delta\phi=0^{\rm
P}\!\!.0863$, and Marsh (1988$^{\cite{Marsh88}}$) obtained
$\Delta\phi=0^{\rm P}\!\!.0858$. The width of the white dwarf
eclipse gives the relation between the orbit inclination $i$ and
the mass ratio $q$ as (see, e.g., Horne, Lanning \& Gomer
1982$^{\cite{Horne82}}$; Dhillon, Marsh \& Jones
1991$^{\cite{Dhillon91}}$):

\begin{equation}
\left(\frac{A\cdot\tan(\pi\Delta\phi)\cdot\sin i}{Y_{RL}}\right)^2
+\left(\frac{A\cdot\cos i}{Z_{RL}}\right)^2=1\,.~~~
\label{eq1}
\end{equation}
Here $Z_{RL}(q)$ and $Y_{RL}(q)$ are the Roche lobe sizes in $Z$
and $Y$ directions ($Z$ axis is perpendicular to the orbital
plane and $Y$ axis is directed along the orbital motion of
mass-losing star). Usually $Z_{RL}(q)$ and $Y_{RL}(q)$ functions
are taken approximately (Pacy\'nski 1971$^{\cite{pacyn71}}$) as:

\begin{equation}
\begin{array}{c}
Z_{RL}(q)/A=Y_{RL}(q)/A=R_{RL}/A\\
~\\
\qquad\qquad
\approx0.462\cdot\left(\dfrac{q}{1+q}\right)^{1/3}\,.
\end{array}
\label{eq2}
\end{equation}
Resulting dependency between $i$ and $q$ for the value
$\Delta\phi=0^{\rm P}\!\!.0863$ is shown in Fig.~1 (dashed
line). However, usage of the approximate formula (\ref{eq2}) may
in this case lead to the loss of accuracy, so we have
calculated dependencies $Z_{RL}(q)$ and $Y_{RL}(q)$ exactly for
$0.3\le q\le 0.7$ and approximated it (error $<0.1\%$) as

\[
Y_{RL}(q)/A\approx-0.14858\cdot q^2+0.32031\cdot q+0.18991\,,
\]
\[
Z_{RL}(q)/A\approx-0.14466\cdot q^2+0.30559\cdot q+0.18320\,.
\]
Obtained relationship $i(q)$ is shown in Fig.~1 as well
(solid line). Besides, we put values $i$ and $q$ adopted by Wood
\& Crawford (1986$^{\cite{WoodCrawford86}}$); Marsh
(1988$^{\cite{Marsh88}}$); and Beekman et al.
(2000$^{\cite{Beekman2000}}$).

Using the values for $K_1$, $K_2$ mentioned above, as well as
the more precise dependency $i(q)$ we take the parameters for IP
Peg as follows:  $M_1=1.02M_\odot$, $M_2=0.5M_\odot$,
$i=82^\circ$.  The distance between the inner Lagrangian point
$L_1$ and accretor is $D=0.573A=0.812R_\odot$, the distance
between system's center of mass and accretor is
$0.329A=0.466R_\odot$.

These parameters give the rotational broadening of the absorbtion
lines of mass-losing star as $V_{rot}=143$ km~s$^{-1}$.
Observational measurements of $V_{rot}$ give $V_{rot}=146$
km~s$^{-1}$ (Harlaftis 1999$^{\cite{Harlaftis99}}$) and
$V_{rot}=125$ km~s$^{-1}$ (Catal\'an, Smith \& Jones
2001$^{\cite{Catalan2000}}$). Good agreement between calculated
and observational parameters in this additional relation bears
witness on the correctness of system's parameters in use.

We would like to stress that there were some factors not
included in our model. Martin et al. (1987$^{\cite{Martin87}}$)
and Beekman et al. (2000$^{\cite{Beekman2000}}$) point out a
possible ellipticity of binary components' orbits (eccentrisity
$e=0.05-0.075$). Some authors (Wood et al.
1989$^{\cite{Wood89}}$; Wolf et al. 1993$^{\cite{Wolf93}}$)
argue that some observational peculiarities of IP Peg can be
explained by the presence of the third body in the system. In our
work we assume that the system contains two stellar components
only, its orbits being circular.

\begin{figure*}[p]
\centerline{\hbox{\psfig{figure=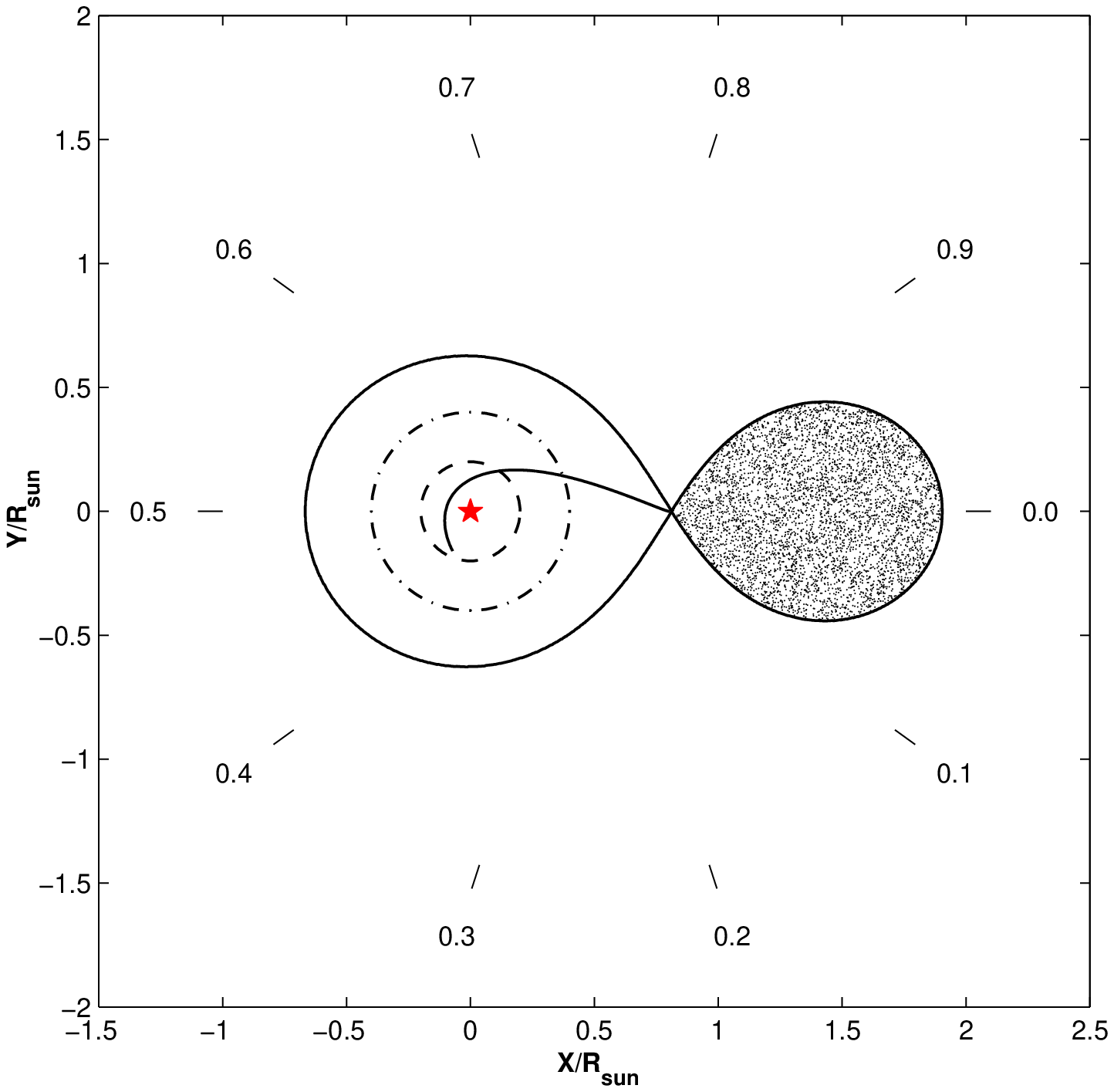,width=10cm}}}
\centerline{\hbox{\psfig{figure=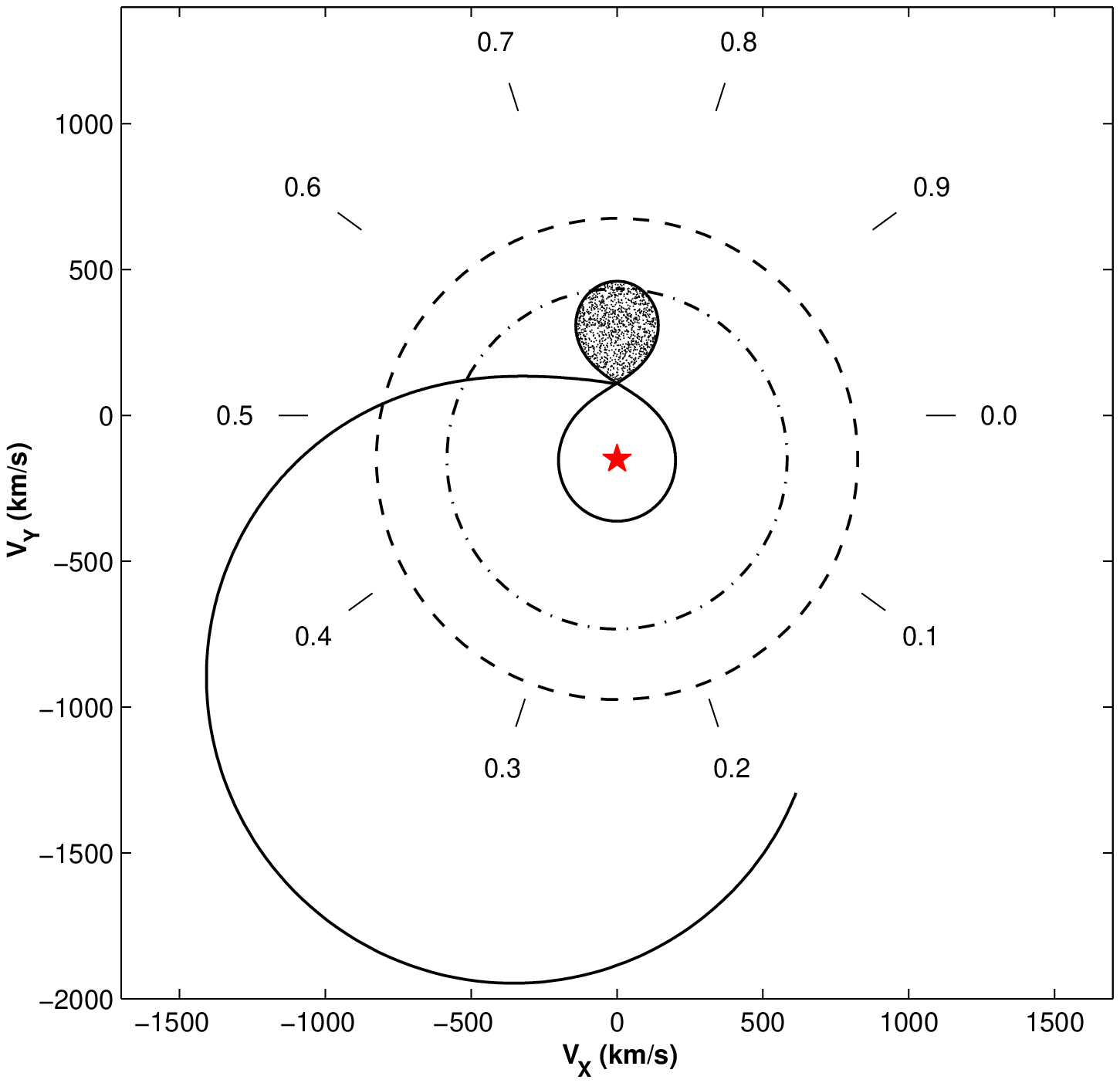,width=10cm}}}
\caption{\footnotesize {\normalsize\it Upper panel:} The adopted
coordinate system with phase angles of observer in a binary
system. The asterisk is the accretor. Orbital rotation of
the binary is counter-clockwise. The Roche lobe filling
donor-star is shadowed. The critical Roche lobe and ballistic
trajectory of a particle moving from $L_1$ are shown by a solid
lines. Dashed and dash-dotted lines show concentric circles
correspond to different radii of the disk.\protect\\[3mm]
{\normalsize\it Lower panel:} The adopted
coordinate system in the velocity space. All designations are
the same as in the upper panel.}
\end{figure*}

\section {Model}

\subsection {Gasdynamical model}

The full description of the 3D gasdynamical model can be found
in Bisikalo et al. (2000$^{\cite{lowvisc}}$). Here we pay
attention only to the main features of the model. To describe
the gas flow in this binary system we used the 3D system of
Euler equations for Cartesian coordinate system. To close the
system of equations, we used the equation of state of ideal gas
with adiabatic index $\gamma$. To mimic the system with
radiative losses, the value of adiabatic
index has been accepted close to unit:  $\gamma = 1.01$, that
corresponds to the case close to the isothermal one (Sawada,
Matsuda \& Hachisu 1986$^{\cite{spiral1}}$, Molteni, Belvedere
\& Lanzafame 1991$^{\cite{diego91}}$; Bisikalo et al.
1995$^{\cite{isoterm}}$).

The calculations were carried out in the non-inertial Cartesian
coordinate system rotating with the binary system. The results
of the calculations and the Doppler tomograms will be presented
in coordinate system defined as follows: the origin of
coordinates is located in the center of the accretor, $X$-axis
is directed along the line connecting the centers of stars, from
accretor to the mass-losing component, $Y$-axis is directed in
the direction of orbital movement of the donor-star, $Z$-axis
is directed along the axis of rotation, so we obtain a
right-hand coordinate system. Figure~2 (top panel) shows the
coordinate system. In this figure we also put digits showing the
phase angles of the observer in binary system, a Roche lobe with
shadowed donor-star and the ballistic trajectory of a particle
moving from $L_1$ point to the accretor. The adopted coordinate
system for Doppler maps is shown in the lower panel of Fig.~2.
The transformation of the donor-star from spatial to velocity
coordinate system is very simple as it is fixed in the
corotating frame. Every point $\bmath r$ fixed in the binary
frame has a velocity ${\bmath\Omega}\times{\bmath r}$ in
the corotation frame. This expression is linear in the
perpendicular distance from the rotation axis, therefore the
shape of the donor-star projected on the orbital plane is
preserved. Since the velocity of each point of the donor-star is
perpendicular to the radius vector, all points of the donor-star
are rotated by 90$^\circ$ counter-clockwise between the spatial
and velocity coordinate diagrams (shadowed regions on the top
and lower panels of Fig.~2, see also Marsh \& Horne
1988$^{\cite{MarshHorne88}}$). On the velocity plane the
accretor has coordinates (0,$K_1$). Figure~2 also depicts two
concentric circles corresponding to different radii of the disk
and their representations on the velocity plane (for Keplerian
rotation law). Inner circle has a larger velocity and forms the
outer circle on the Doppler tomogram.

To obtain numerical solution of the system of equations we used
the Roe--Osher TVD scheme of a high approximation order (Roe
1986$^{\cite{Roe}}$; Chakravarthy \& Osher
1985$^{\cite{Osher}}$) with Einfeldt modification (Einfeldt
1988$^{\cite{Einfeldt}}$). The computational domain was taken as
a parallepipedon $[-D\ldots D]\times[D\ldots D]\times[0\ldots
\slantfrac{1}{2}D]$ (due to the symmetry of the problem
calculations were conducted only in the top half-space). A
sphere with a radius of $\slantfrac{1}{100}A$ representing the
accretor was cut out of the calculation domain. The boundary
conditions were taken as `free outflow' on the accretor star and
on the outer edges of computational domain. In the gridpoint
corresponding to $L_1$ we injected the matter with parameters
$\rho=\rho(L_1)$, $V_x=c(L_1)$, $V_y=V_z=0$, where $c(L_1)$ is a
gas speed of sound in $L_1$ point. Due to the scaling of the
system of equations with respect to $\rho$ (with simultaneous
scaling of pressure $P$) we can accept an arbitrary value of
$\rho(L_1)$ so we take it as $\rho(L_1)=1$.\footnote{When
considering a system with known mass-loss rate, to determine the
real values it is necessary to change the calculated values of
density in accordance with the scale, defined by the ratio of
the real value of the mass loss rate to the model one.} The
sound speed in $L_1$ was adopted as 5.5~km~s$^{-1}$ which
corresponds to $T(L_1)=3500$~K. For the initial conditions we
used rarefied gas with the following parameters
$\rho_0=10^{-5}\cdot\rho(L_1)$,
$P_0=10^{-4}\rho(L_1)c^2(L_1)/\gamma$, ${\bmath V}_0=0$.

\renewcommand{\thefigure}{3a}
\begin{figure*}[t]
\centerline{\hbox{\psfig{figure=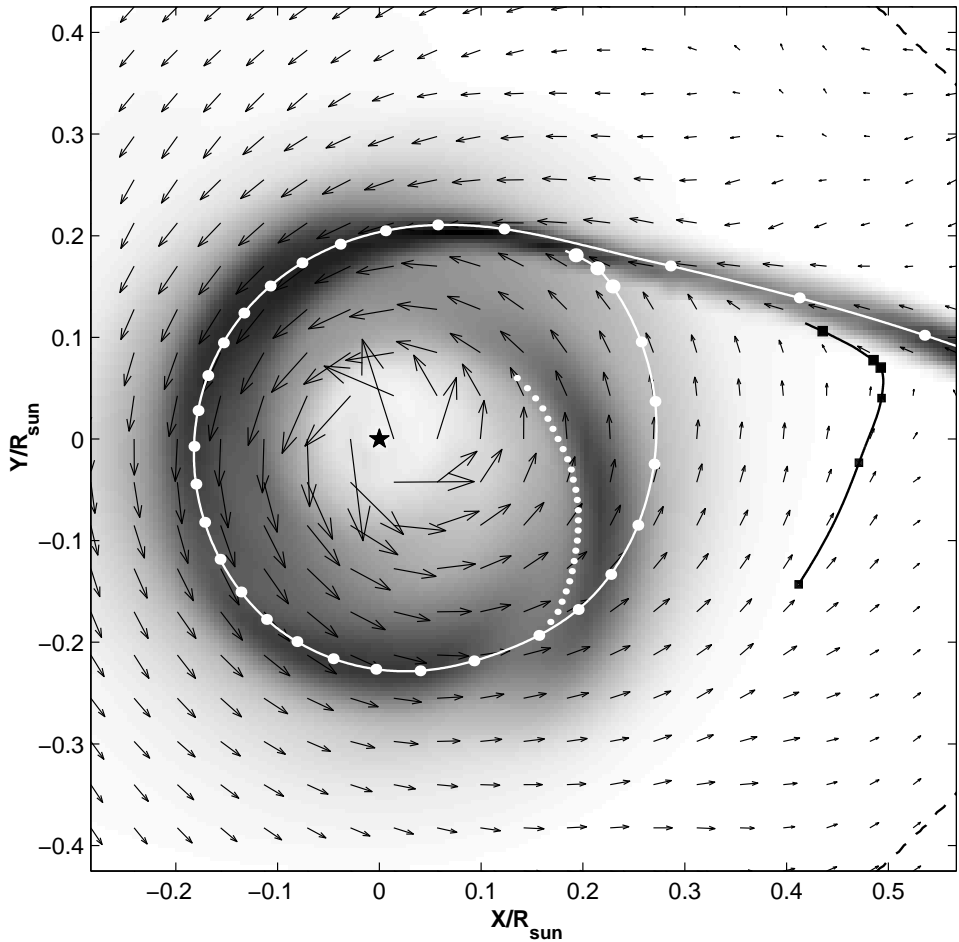,width=13cm}}}
\caption{\footnotesize The distribution of density over the
equatorial plane. Arrows are the velocity vectors in
observer's frame. The asterisk is the accretor. The
dashed-dotted line is Roche equipotential passing through $L_1$.
The white dotted line is the tidally induced spiral shock.
Gasdynamical trajectory of a particle moving from $L_1$ to
the accretor is shown by a white line with circles. Another gas
dynamical trajectory is shown by a black line with squares (see
also Fig.~4).}
\end{figure*}

\renewcommand{\thefigure}{3b}
\begin{figure*}[t]
\centerline{\hbox{\psfig{figure=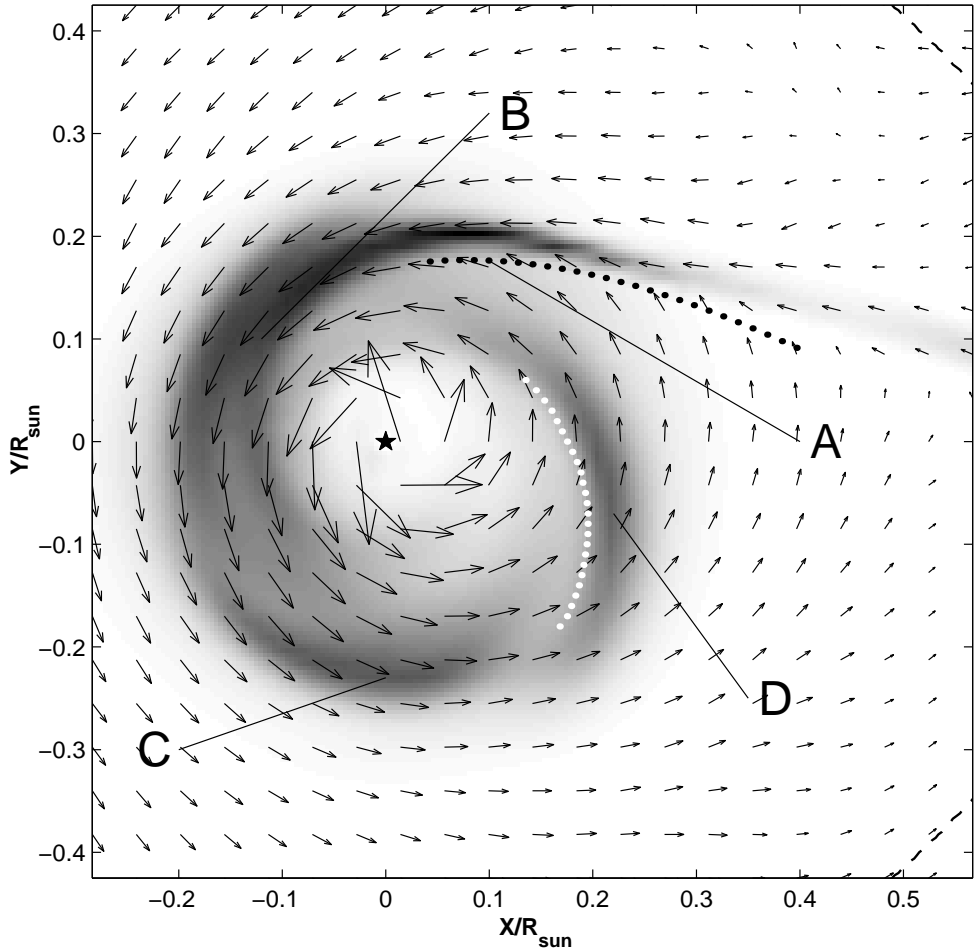,width=13cm}}}
\caption{\footnotesize The distribution of $\rho^2 T^{1/2}$ over
the equatorial plane. A black dotted line is the shock wave
along the edge of the stream (`hot line'). The main emission
regions are marked by {\bf A}, {\bf B}, {\bf C}, {\bf D}. Other
designations are the same as in Fig.~3a.}
\end{figure*}

\renewcommand{\thefigure}{4}
\begin{figure*}[t]
\centerline{\hbox{\psfig{figure=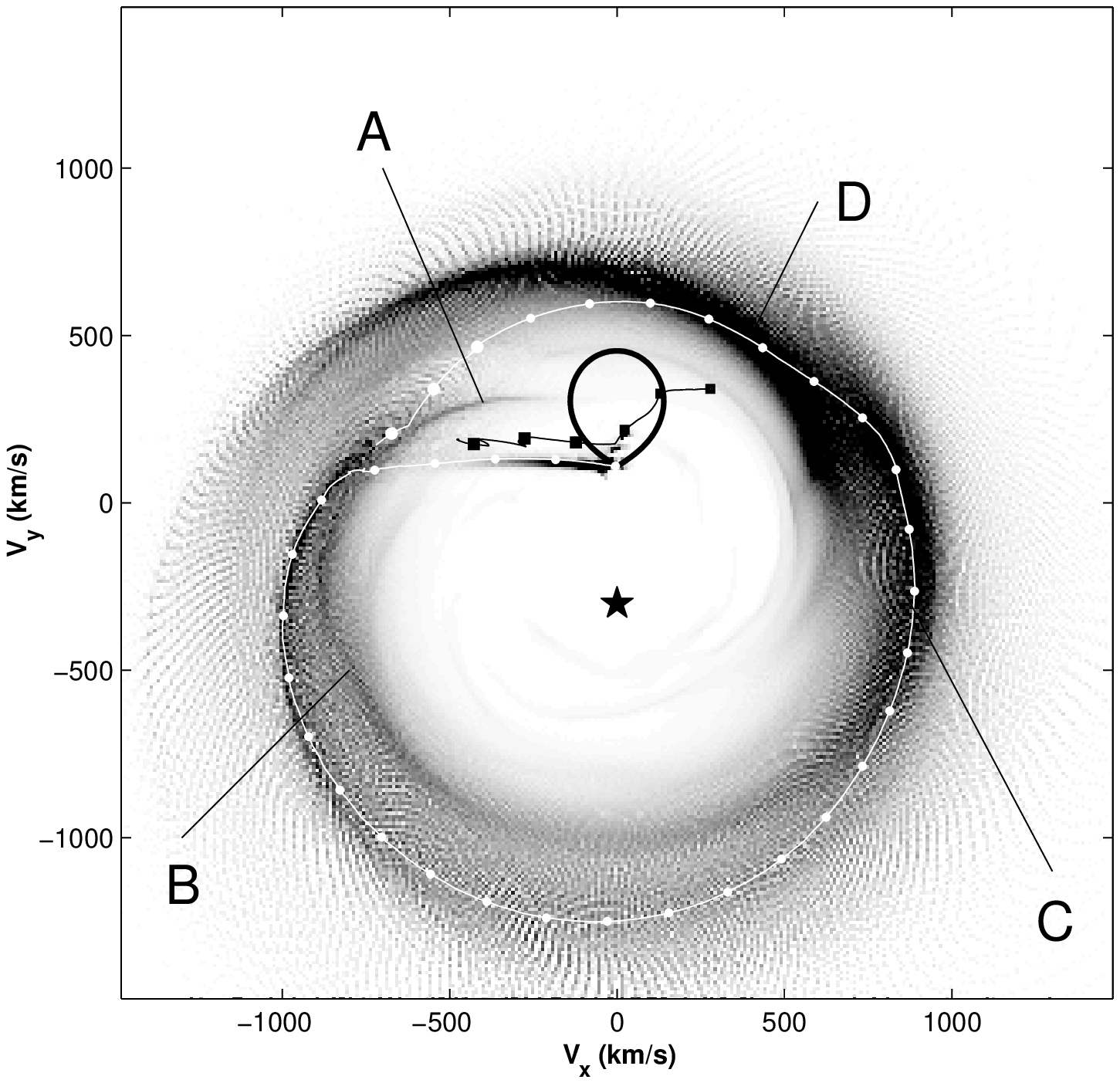,width=13cm}}}
\caption{\footnotesize Synthetic Doppler map for
$I\sim\rho^2T^{1/2}$. The secondary Roche lobe (a bold black
line) and the accretor (an asterisk) are also shown. The
white line with circles and black line with squares show gas
dynamical trajectories in the velocity coordinates (see
Fig.~3a). The main emission regions are marked by {\bf A}, {\bf
B}, {\bf C}, {\bf D} as in Fig.~3b.}
\end{figure*}

\renewcommand{\thefigure}{5}
\begin{figure*}[t]
\centerline{\hbox{\psfig{figure=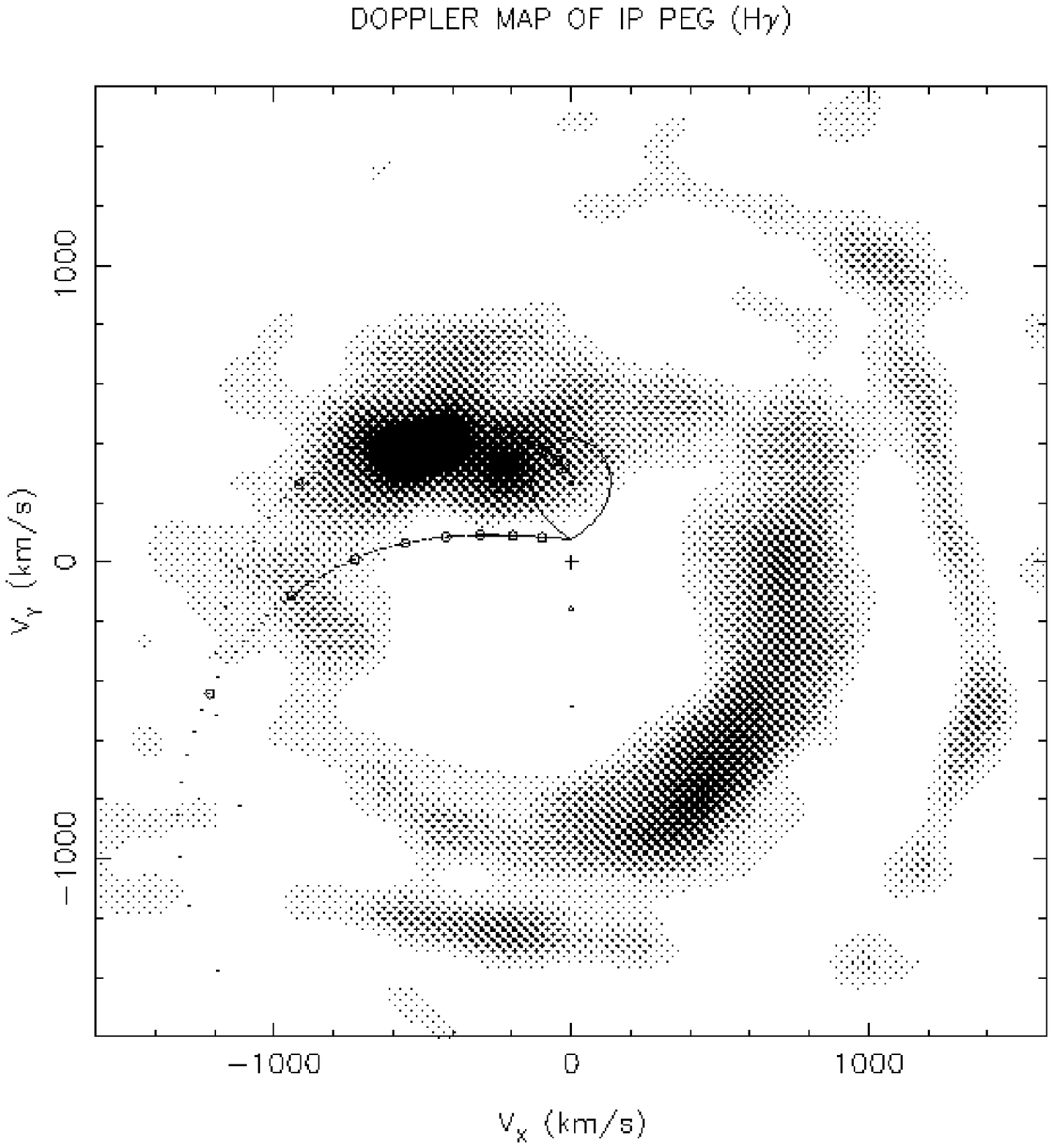,width=10cm}}
\hbox{\psfig{figure=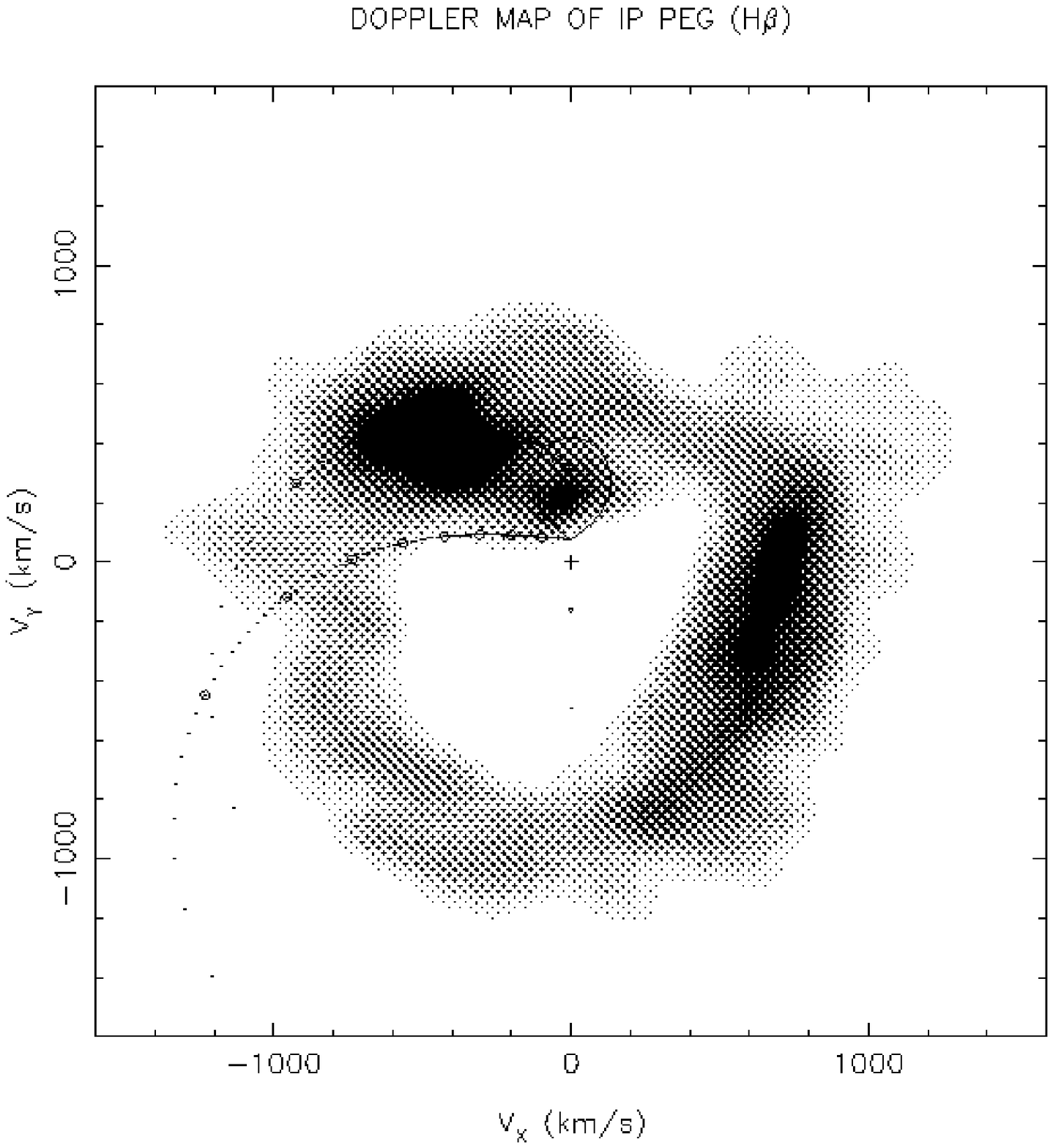,width=10cm}}}
\caption{\footnotesize Doppler maps of H$_\gamma$ and
H$_\beta$ lines for IP Peg in quiescence (Wolf et al.
1998$^{\protect\cite{Wolf98}}$). This figure is reproduced under
the kind permission by S.Wolf and A.Bobinger.}
\end{figure*}

\renewcommand{\thefigure}{6a}
\begin{figure*}[t]
\centerline{\hbox{\psfig{figure=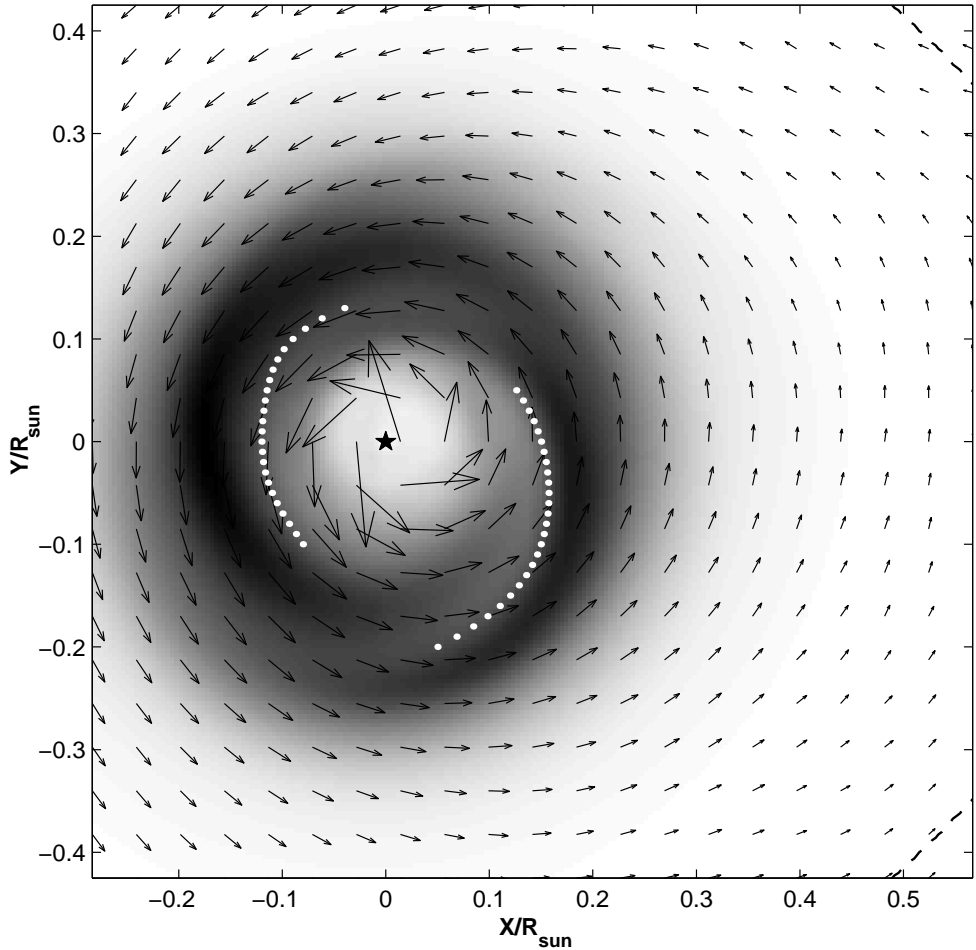,width=13cm}}}
\caption{\footnotesize The distribution of density over the
equatorial plane for outburst. Arrows are the velocity vectors in
observer's frame. The asterisk is the accretor. The
dashed-dotted line is Roche equipotential passing through $L_1$.
White dotted lines are the arms of tidally induced spiral
shock.}
\end{figure*}

\renewcommand{\thefigure}{6b}
\begin{figure*}[t]
\centerline{\hbox{\psfig{figure=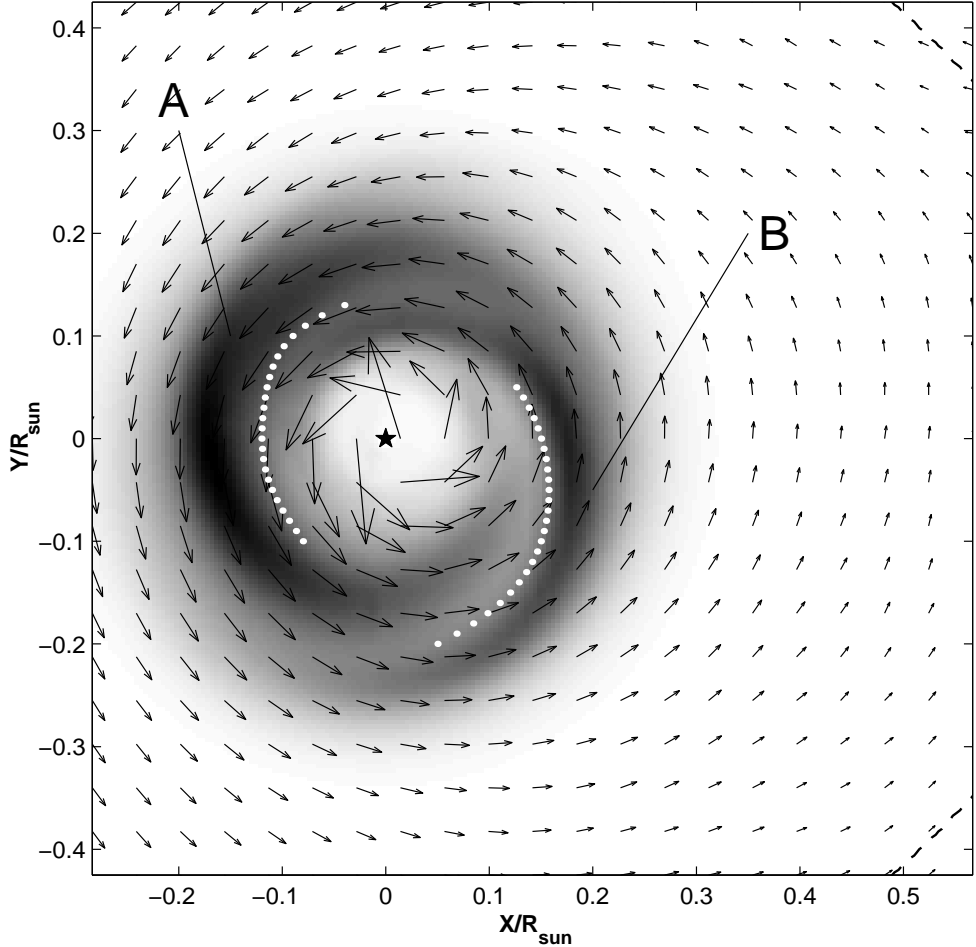,width=13cm}}}
\caption{\footnotesize The distribution of $\rho^2 T^{1/2}$ over
the equatorial plane for outburst. The main emission regions are
marked by {\bf A} and {\bf B}. Other designations are the same
as in Fig.~6a.}
\end{figure*}

\renewcommand{\thefigure}{7}
\begin{figure*}[t]
\centerline{\hbox{\psfig{figure=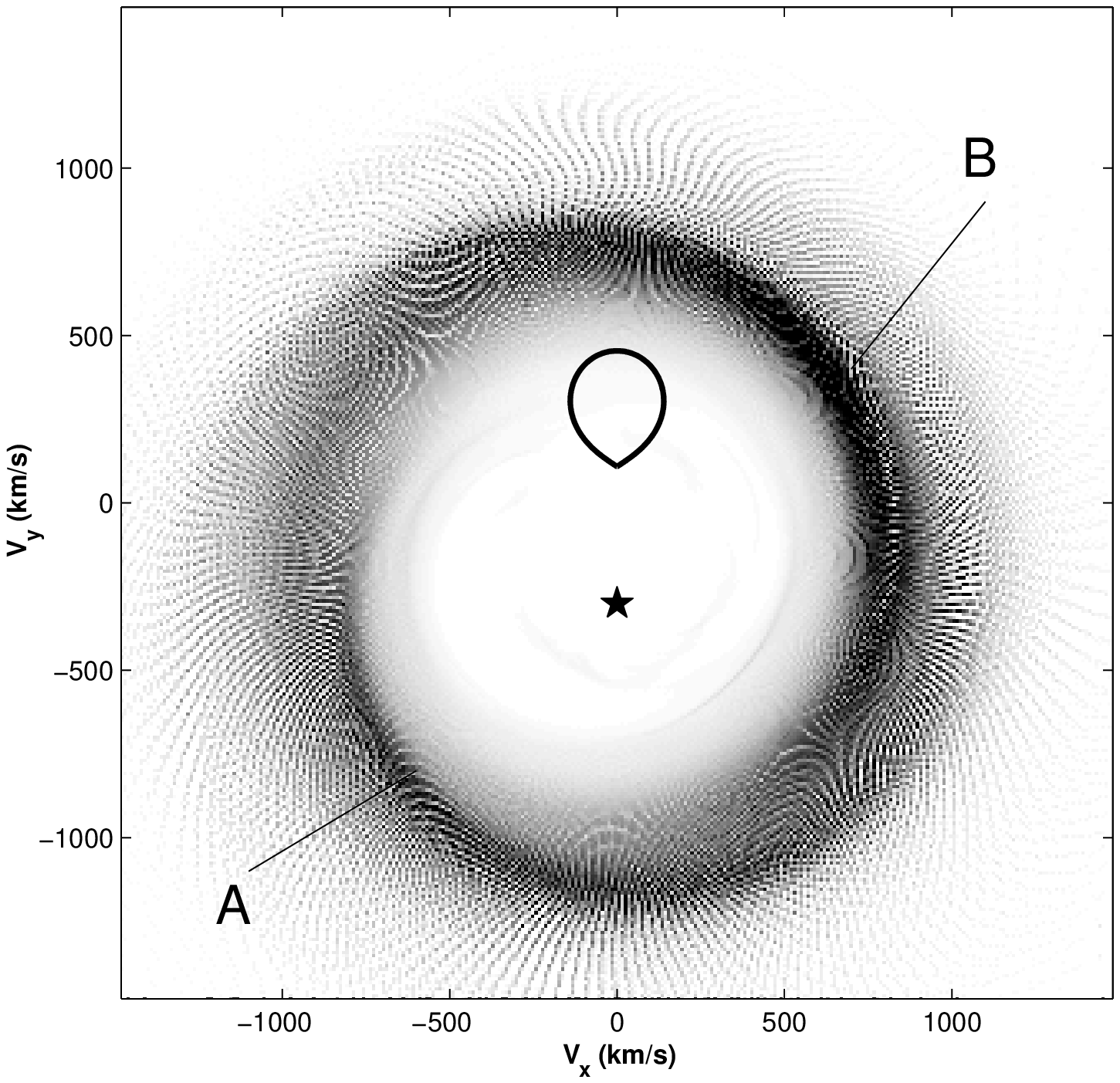,width=13cm}}}
\caption{\footnotesize Synthetic Doppler map for the outburst.
The secondary Roche lobe (a bold black line) and the accretor
(an asterisk) are also shown.  The main emission regions are
marked by {\bf A} and {\bf B} as in Fig.~3b.} \end{figure*}

\subsection {Technique for construction of synthetic Doppler
tomograms}

Preceding the consideration of the synthetic Doppler tomograms
for binary system IP Peg it is necessary to stress the
complexity of the analysis of these tomograms for eclipsing
systems (see, e.g., Kaitchuck et al.
1994$^{\cite{Kaitchuck94}}$).  The concept of Doppler tomography
implies that any geometrical point deposits in some place on the
velocity plane, this point being visible for any orbital phase.
Indeed, from (\ref{eq_appendix}) we have

\[
I(V_R,\phi+\pi)=I(-V_R,\phi)\,,
\]
i.e. the constructions of synthetic Doppler tomograms is
possible only for those sets of trailed spectrograms
$I(V_R,\phi)$ which are transformed symmetrically when one sees
the binary system from the \lqq dark side" or, in other words, when
there are no eclipses and occultations of emission regions.
Clearly, eclipsing systems violates that assumption. Usually,
when dealing with observations, the eclipsed parts of trailed
spectrograms are naturally excluded from input data for
construction of Doppler tomograms. As a result, we obtain the
Doppler map corresponding to the \lqq transparent" case. The
conversion of the results of gasdynamical simulation into
Doppler maps suggests using of full set of data that also
corresponds to the \lqq transparent" case.

The Doppler maps show the distribution of luminosity in
the velocity space. Each point of the flow has a
three-dimensional vector of velocity ${\bmath U}=(U_x,U_y,U_z)$
in observer's (inertial) frame. In the case when observer is
located in the orbital plane of the binary the Doppler map's
coordinate $(V_x,V_y)$ will coincide with $U_x$ and $U_y$. To
define these coordinates for the case of inclined system we have
to find a projection of vector ${\bmath U}$ on the plane
constituted by vectors $\bmath n$ and ${\bmath
n}\times{\bmath\Omega}$, where $\bmath n$ is a direction from
the observer to binary.

The line emissivity in the velocity space can be written as:

\begin{equation}
\begin{array}{l}
I(V_x,V_y)\sim\displaystyle\int\limits_{\cal O}
\int\limits_{U_x}\int\limits_{U_y} I(x,y,z)\\
~\\
\qquad\times
\delta(U_x(x,y,z)\sin i+U_z(x,y,z)\cos i-V_x)\\
~\\
\qquad\times
\delta(U_y(x,y,z)\sin i+U_z(x,y,z)\cos i-V_y)\\
~\\
\qquad\times
d{\cal O}dU_xdU_y\,,
\end{array}
\label{eq3}
\end{equation}
where $d{\cal O}=dxdydz$, $i$ -- inclination angle.

As was mentioned above we adopt intensity as
$I\sim\rho^2T^{1/2}$ for the construction of synthetic Doppler
tomograms.

\section{Results for quiescence of IP Peg}

Based on the model described in the Section 4.1 we have
conducted the 3D gasdynamical simulation of IP Peg in quiescence
up to reaching of a steady-state solution. The morphology of
gaseous flows in considered binary system can be evaluated from
Figs~3a. In Fig.~3a the distribution of density over the
equatorial plane and velocity vectors are presented. In this
Figure we also put a gasdynamical trajectory of a particle
moving from $L_1$ to accretor (a white line with circles) and a
gasdynamical trajectory passing through the shock wave along
the stream edge (a black line with squares, see also Fig.~4).
Analysis of the presented results as well as our previous studies
(Bisikalo et al. 1997$^{\cite{paper1}}$, 1998b$^{\cite{mnras}}$)
shows the significant influence of the rarefied gas of
circumbinary envelope on the flow patterns in semidetached
binaries. The gas of circumbinary envelope interacts with the
stream of matter and deflects it. This leads, in particular, to
the shock-free (tangential) interaction between the stream and
the outer edge of forming accretion disc, and, as the
consequence, to the absence of `hot spot' in the disc.

At the same time it is seen, that the interaction of the gas of
circumbinary envelope with the stream results in the formation
of an extended shock wave located along the stream edge (`hot
line'). The `hot line' model was confirmed by confronting with
observations (see Bisikalo et al. 1998a$^{\cite{Tanya98}}$;
Khruzina et al. 2001$^{\cite{Tanya2001}}$). From Fig.~3a it is
also seen the formation of tidally induced spiral shock
(white dotted line in Fig.~3a). Appearance of the tidally
induced two-armed spiral shock was numerically discovered in
Sawada, Matsuda \& Hachisu (1986a$^{\cite{spiral1}}$,
1986b$^{\cite{spiral2}}$); Sawada et al.
(1987$^{\cite{Sawada87}}$); Spruit et al.
(1987$^{\cite{Spruit98}}$); Matsuda et al.
(1990$^{\cite{Matsuda90}}$). Here we see only the one-armed
spiral shock. In the place where the second arm should be the
stream from $L_1$ dominates and presumably prevents the
formation of second arm of tidally induced spiral shock.

An analysis of the flow structure out of equatorial plane
shows that a part of the circumbinary envelope
interacts with the (denser) gas stream and overflows it. This
naturally leads to the formation of `halo'. Following to
(Bisikalo et al. 2000$^{\cite{lowvisc}}$), one can define `halo'
as that matter which: i) encircles the accretor being
gravitationally captured; ii) does not belong to the accretion
disc; iii) interacts with the stream (collides with it and/or
overflows it); iv) after the interaction either becomes a part
of the accretion disc or leaves the system.

Figure~3b depicts the distribution of $\rho^2T^{1/2}$ over the
equatorial plane. Similar to Fig.~3a spiral shock is shown by
white dotted line. Besides, shock wave along the edge of the
stream is shown by black dotted line. The distribution shown in
Fig.~3b represents the intensity of recombination line, so the
analysis of this distribution can determine the most luminous
region of the flow. It is seen, that the main emission comes
from four region designated by markers {\bf A}, {\bf B}, {\bf C},
{\bf D}.

\begin{itemize}

\item  Marker {\bf A} designates the shock wave along the edge of
the stream (`hot line') resulting from the gasdynamical
interaction of the gas of circumbinary envelope with the stream.

\item Marker {\bf B} designates the stream from $L_1$ or, more
exactly, the most luminous part of the stream where the density
is still large enough and the temperature already increases due
to dissipation.

\item Marker {\bf C} designates a region near the apoastron of
the accretion disk. The analysis of the presented results shows
that the disk has a quasi-elliptical form, therefore approaching
the apoastron the matter is retarded and the dense region is
formed.

\item Marker {\bf D} designates a dense post-shock region
attached to the spiral shock.

\end{itemize}

\noindent
The synthetic Doppler map based on the results of 3D
gasdynamical simulations is presented in Fig.~4. Earlier we have
analyzed the features of flow in the equatorial plane (see
Figs~3a,~3b), but for the sake of comparison with observations
synthetic Doppler maps will be presented for integrated over
$z$-coordinate intensity in accordance to the equation
(\ref{eq3}).  Two gasdynamical trajectories (the same as in
Fig.~3a but in velocity coordinates) are shown in Fig.~4.

Shock wave resulting from the gasdynamical interaction of the
gas of circumbinary envelope with the stream is located along
the edge of the stream. Three last points (marked by larger
symbols) of curves with circles and squares in Fig.~3a are the
examples of two flowlines passing through the shock. Location of
these parts of trajectories on the Doppler map corresponds to
region {\bf A} left to the donor-star. This region of Doppler
map contains also a spiral arm beginning approximately from the
center of mass-losing star and located above the vicinity of
$L_1$. Our analysis shows that the appearance of this region
results from the overflowing the stream by the gas of
circumbinary envelope but not from the shock wave. It is seen
from Fig.~4 that the stream from $L_1$ (the beginning of white
line whit circles) transforms into spiral arm {\bf B} in III
quadrant\footnote{The quadrants of coordinates plane are counted
as follows: I quadrant is upper right (corresponding to $V_x>0$,
$V_y>0$), other quadrants are counted counter-clockwise.} of the
Doppler map. The region of increased density near apoastron of
the disk -- region {\bf C} is seen in Fig.~4 as more luminous
zone on the border of I and IV quadrants. Tidally induced spiral
shock (or, more exactly, the dense post-shock zone, dotted line
in Figs~3a,~3b) forms a bright arm in I and II quadrants of
Doppler map.

Resuming these results for Doppler map of IP Peg in quiescence
we can conclude that there are four elements of the flow
structure which deposit in the total luminosity:  `hot line',
the most luminous part of the stream where, the dense region
near the apoastron of the disk, and the dense post-shock region
attached to the spiral shock. The income of each element
obviously can vary depending on peculiarities of considered
binary system.  It is also obvious that based on the model
computations we can't estimate what elements will dominate.
Nevertheless, comparison of synthetic Doppler maps and observed
ones permits both to catch the dominating element and to
correct/refine the computational model.

Observational Doppler tomograms for IP Peg in quiescence were
built in Marsh \& Horne (1990$^{\cite{MarshHorne90}}$); Harlaftis
et al. (1994$^{\cite{Harlaftis94}}$); Wolf et al.
(1998$^{\cite{Wolf98}}$); Bobinger et al.
(1999$^{\cite{Bobinger99}}$); Bobinger
(2000$^{\cite{Bobinger2000}}$). Figure~5 represents a typical
Doppler map for H$_\gamma$ and H$_\beta$ from Wolf et al.
(1998$^{\cite{Wolf98}}$). The characteristic features of these
tomograms are the bright spot in the region {\bf A}
as well as the zone of moderate brightness in the region {\bf
C}. The comparison of the observational tomogram from Fig.~5 and
synthetic one from Fig.~4 reveals that it is the `hot line' and
the dense zone near the disk's apoastron which mainly deposit into
the total luminosity. Signatures of the spiral shock are not
seen and this implies either its absence or weakness. Note also
that the observational tomogram shows rather small input from
the stream from $L_1$ into the total luminosity.

\section{Results for outburst of IP Peg}

Observations show (see, e.g., recent reviews by  Marsh
2000$^{\cite{Marsh2000}}$ and Steeghs
2000$^{\cite{Steeghs2000}}$) that during outburst the accretion
disk dominates hence the stream from $L_1$ plays less important
role. Our today's knowledge of the nature of the outburst as
well as its parameters has an approximate and qualitative
character so it is hard to simulate the outburst correctly. To
mimic the flow structure during outburst on the qualitative
level we calculated the structure of gaseous flows up to
reaching a quasi-steady-state solution and put the rate of mass
transfer equal to zero (i.e. terminated the mass transfer) as it
was suggested in Bisikalo et al. (2001a$^{\cite{blob}}$,
2001b$^{\cite{blob2}}$). We understand that this model doesn't
reflect all peculiarities of outburst and expanding accretion
disk but we hope that it truly correlates the influence the disk
and the stream on the qualitative level.  Our simulations of
residual accretion disk show that at time $0.3\div0.4P_{orb}$
after mass transfer termination the flow structure is changed
significantly. The stream from $L_1$ vanishes and doesn't
dominate anymore, and the shape of accretion disk changes from
quasi-elliptical to circular. The second arm of tidally induced
spiral shock is formed while earlier (before the termination of
mass transfer) it was suppressed by the stream from $L_1$. It is
seen that obtained flow structure has all basic features
observed in outburst of IP Peg. This gives a hope that we can
refine/reveal the new features of IP Peg in outburst by virtue
of analysis of synthetic Doppler tomograms constructed for this
gasdynamical solution.

Figure~6a depicts the distribution of density and velocity
vectors and Fig.~6b depicts the distribution of $\rho^2T^{1/2}$
over the equatorial plane.  It is seen, that the main emission
comes from two arms of the spiral shock designated by markers
{\bf A} and {\bf B}. The synthetic Doppler map based on the
results of 3D gasdynamical simulations for IP Peg in outburst is
presented in Fig.~7. Our analysis shows that bright arms in I
and III quadrants are due to emission of dense post-shock zones
attached to the arms of spiral shock.

Observational Doppler tomograms for IP Peg during the outburst
were built in Marsh \& Horne (1990$^{\cite{MarshHorne90}}$);
Steeghs et al. (1996$^{\cite{Steeghs96}}$); Steeghs, Harlaftis \&
Horne (1997$^{\cite{Steeghs97}}$); Harlaftis et al.
(1999$^{\cite{HarlaftisEtAl99}}$); Morales-Rueda, Marsh \&
Billington (2000$^{\cite{Luisa2000}}$). A typical example of such
tomogram (Morales-Rueda, Marsh \& Billington
2000$^{\cite{Luisa2000}}$) is given in Fig.~8. The
characteristic features of these tomograms are two bright arms
in I and III quadrants. The comparison of the observational
tomogram from Fig.~8 and synthetic one from Fig.~7 reveals that
these arms results from dense post-shock zones attached to the
arms of spiral shock (zones {\bf A} and {\bf B}).

\section{Conclusions}

Using of the gasdynamical calculations alongside with Doppler
tomography technique permits us to identify main features of the
flow on the Doppler maps without solution of the ill-posed
inverse problem. The comparison of synthetic Doppler maps and
observed ones permits to correct/refine the computational model
and to interpret the observational data. In this work we have
presented the synthetic Doppler maps of gaseous flows in binary
IP Peg based on the results of 3D gasdynamical simulations. The
zones of flow structure responsible for the most emitive regions
of Doppler map were identified and it was found that they are
different for quiescence and outburst. Our analysis for
quiescence has shown that it is the shock wave along the stream
edge -- `hot line' and the dense zone near the disk's apoastron
which mainly deposit into the total luminosity. The input from
the stream from $L_1$ and the spiral shock into the total
luminosity is small. During the outburst the role of the stream
is unimportant and the accretion disk with two-armed spiral
shock dominates. The comparison of the observational tomogram
and synthetic one reveals that the bright arms in the Doppler map
result from dense post-shock zones attached to the arms of
the spiral shock.

\section*{Acknowledgments}

The work was partially supported by Russian Foundation for
Basic Research (projects NN 99-02-17619, 99-02-17589,
00-01-00392, 00-02-16471, 00-02-17253), by grants of President
of Russia (99-15-96022, 00-15-96722, 00-15-96553), and by INTAS
(grant 01-491). Authors wish to thank V.V.Neustroev for useful
discussions on observational Doppler tomograms.

\clearpage

\onecolumn

\renewcommand{\thefigure}{8}
\begin{figure*}[t]
\centerline{\hbox{\psfig{figure=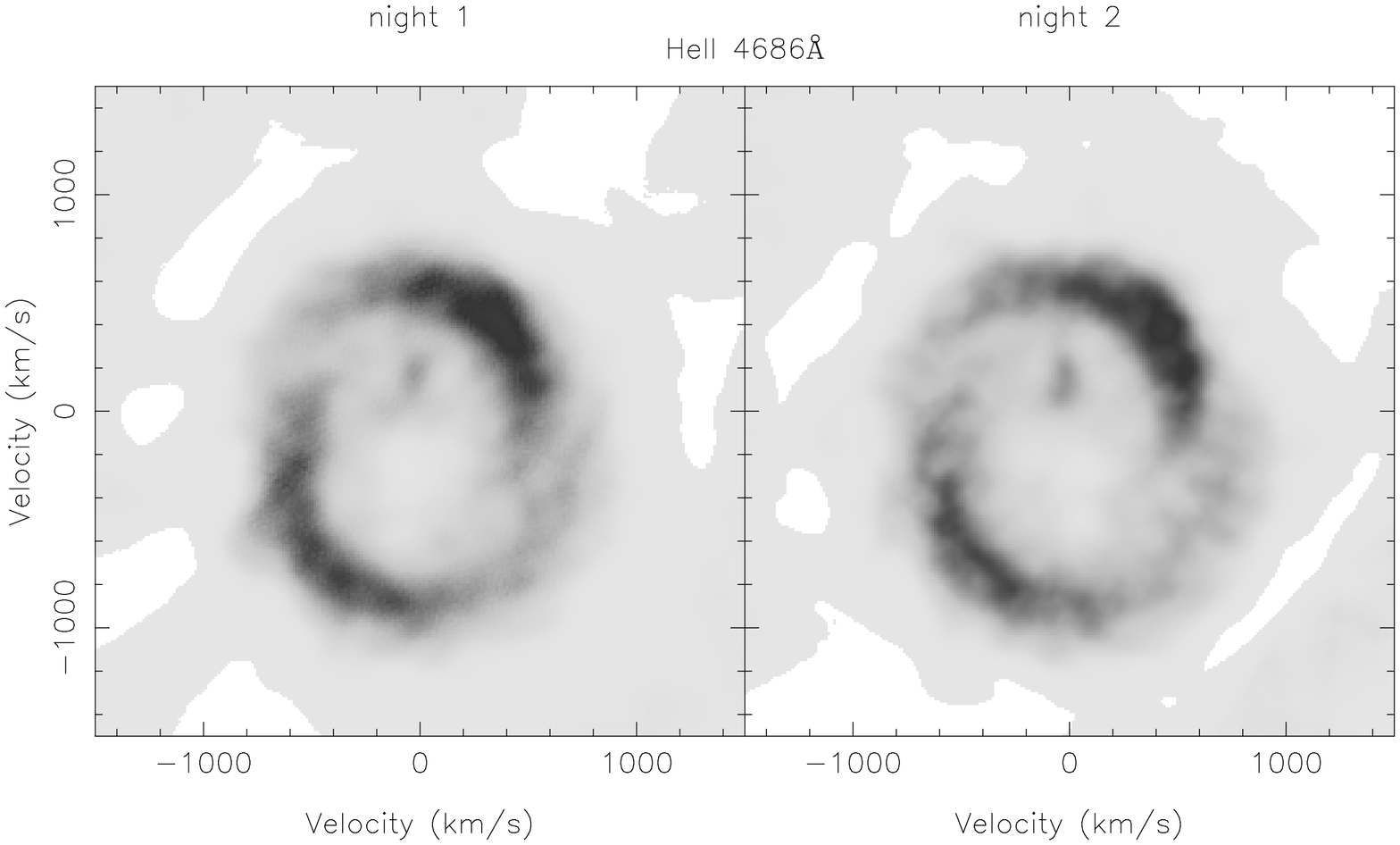,angle=-90,width=16cm}}}
\caption{\footnotesize Doppler maps of He~II~$\lambda~4686$\AA~
for IP Peg in outburst (Morales-Rueda, Marsh \& Billington
1998$^{\protect\cite{Luisa2000}}$). This figure is reproduced
under the kind permission by L.Morales-Rueda.}
\end{figure*}

\end{document}